\documentclass[prd,aps,nofootinbib,superscriptaddress]{revtex4}
\usepackage{epsfig}
\usepackage{amsmath, slashed}
\usepackage{xcolor}
\usepackage{appendix}
\usepackage{dsfont}

\usepackage[normalem]{ulem}

\usepackage{bm}
\usepackage{amsfonts}

\definecolor{pine}{rgb}{0.3, 0.5, 0.3}

\allowdisplaybreaks
\begin{document}

\title{Kinematical higher-twist corrections in $\gamma^* \gamma \to M \bar M $}
              
\author{C\'edric Lorc\'e}
\email[]{cedric.lorce@polytechnique.edu}
\affiliation{CPHT, CNRS, Ecole Polytechnique, Institut Polytechnique de Paris, Route de Saclay, 91128 Palaiseau, France}

\author{Bernard Pire}
\email[]{bernard.pire@polytechnique.edu}
\affiliation{CPHT, CNRS, Ecole Polytechnique, Institut Polytechnique de Paris, Route de Saclay, 91128 Palaiseau, France}

\author{Qin-Tao Song}
\email[]{songqintao@zzu.edu.cn}
\affiliation{CPHT, CNRS, Ecole Polytechnique, Institut Polytechnique de Paris, Route de Saclay, 91128 Palaiseau, France}
\affiliation{School of Physics and Microelectronics, Zhengzhou University, Zhengzhou, Henan 450001, China}


\date{\today}

\begin{abstract}
We estimate kinematical higher-twist (up to twist 4) corrections to the $\gamma^*(q_1) \gamma(q_2) \to M(p_1) \bar{M}(p_2)$ amplitudes at large $Q^2=-q_1^2$ and small $s=(q_1+q_2)^2$, where $M$ is a scalar or pseudoscalar meson. This process is known to factorize at leading twist into a perturbatively calculable coefficient function and generalized distribution amplitudes (GDAs). The kinematical higher-twist contributions of order $s/Q^2$ and $m^2/Q^2$ turn out to be important in the cross section, considering the kinematics accessible at Belle  and Belle II.
 We  present  numerical estimates for the cross section for $\gamma^* \gamma \to \pi^0 \pi^0$ with the $\pi \pi$ GDA extracted from Belle measurements and with the asymptotic $\pi \pi$ GDA as inputs to study the magnitude of the kinematical corrections. 
To see how the target mass corrections of order $m^2/Q^2$ affect the cross section, we also perform the calculation for $\gamma^* \gamma \to \eta \eta$ by using a model
$\eta \eta$ GDA.
In the range   $s> 1$ GeV$^2$, 
the kinematical higher-twist corrections account for $\sim 15 \%$ of the total cross section, an effect which is not negligible.
Since $\pi \pi$ GDAs are the best way to access the pion energy-momentum tensor (EMT), our study demonstrates that an accurate evaluation of EMT form factors requires the inclusion of kinematical higher-twist contributions.

\end{abstract}

\maketitle

\section{Introduction}

Generalized distribution amplitudes (GDAs) \cite{Muller:1994ses, Diehl:1998dk,Polyakov:1998ze} -- sometimes called two-meson distribution amplitudes -- are hadronic matrix elements closely related to generalized parton distributions (GPDs) \cite{Diehl:2003ny,Belitsky:2005qn, Boffi:2007yc, Goeke:2001tz}. They involve the same bilocal quark (or gluon) operator on the light cone and correspond to $s$-$t$ crossed helicity matrix elements. GDAs can be accessed in $e(k_1) \gamma \to e' (k_2) M(p_1) \bar{M}(p_2)$ reactions in $e^+ e^-$ collisions, in the kinematical range where $Q^2= -(k_1-k_2)^2$ is large but $s = (p_1+p_2)^2$ is much smaller than $Q^2$. They have already been the subject of careful studies at Belle \cite{Belle:2015oin} and were extracted in a leading-twist analysis in Ref.\,\cite{Kumano:2017lhr}. They are also important in the understanding of heavy meson three-body decays, in particular in the quest for a precise determination of Cabibbo-Kobayashi-Maskawa (CKM) matrix elements \cite{Chen:2002th,Wang:2015uea, Li:2016tpn, Jia:2021uhi}. 

As the studies of GPDs allow us to perform nucleon tomography through a Fourier transform in the transverse coordinate space \cite{Burkardt:2000za, Ralston:2001xs, Diehl:2002he}, GDAs open the way to an impact-parameter picture \cite{Pire:2002ut} of the exclusive hadronization process $q \bar q \to M(p_1) \bar{M}(p_2)$.
The GPDs and GDAs are also used to investigate the matrix elements of the energy-momentum tensor (EMT) \cite{Ji:1996ek,Ji:1996nm, Diehl:2000uv, Polyakov:1998ze} for hadrons in the spacelike
and timelike regions, respectively.
 One can extract mass, pressure and shear force distributions of hadrons with the spacelike EMT form factors \cite{Polyakov:2002yz, Goeke:2007fp, Mai:2012cx, Polyakov:2018zvc,Burkert:2018bqq,  Kumericki:2019ddg,Lorce:2018egm, Dutrieux:2021nlz}.
Since there is no experimental facility where pion GPDs can be directly measured (see however Refs.\,\cite{Amrath:2008vx, Chavez:2021koz, Chavez:2021llq}), the studies of $\pi \pi$ GDAs are a necessary tool to access the pion EMT.
The spacelike pion EMT form factors  can be obtained from the timelike ones by using dispersion relations,
and in this process the $\pi \pi$ GDAs and pion EMT form factors at $s> 1$ GeV$^2$ should be included so as to make the integrals convergent. Therefore, 
this goal  necessitates to extract GDAs in a sufficiently large $s$-range, thus demanding a control as precise as possible of kinematical higher-twist corrections to the amplitudes, which are proportional to $s/Q^2$ and $m^2/Q^2$, with $m$ the meson mass. While the contribution of one higher-twist process has been previously discussed in Ref.\,\cite{Lansberg:2006fv}, it only matters in a very limited kinematical region, namely the near forward or near backward regions. The phenomenological necessity of a sizeable (genuine) twist-$4$ contribution to the $\gamma^*\gamma \to \rho \rho$ amplitude has also been pointed out \cite{Anikin:2005ur}. A complete understanding of higher-twist corrections to the process 
\begin{align}
\gamma^*(q_1) \gamma(q_2) \to M(p_1) \bar{M}(p_2)\,, 
\label{process}
\end{align}
is however a difficult task which is far from being achieved.

Meanwhile, a separation of kinematical and dynamical contributions in the product of two electromagnetic currents
$T \{j_{\mu}^\text{em}(z_1x)j_\nu^\text{em} (z_2x) \}$  was proven in Refs.\,\cite{Braun:2011dg,Braun:2011zr,Braun:2011th} and applied to the deeply-virtual Compton scattering (DVCS) reaction \cite{Braun:2012bg}. 
The kinematical corrections come from two types of operators, namely 
the subtraction of traces in the leading-twist operators and the higher-twist operators which can be reduced to the total derivatives of the leading-twist ones. The subtraction of traces was applied in Ref.\,\cite{Nachtmann:1973mr} to the reaction of Deep Inelastic Scattering (DIS), leading to target mass corrections. 
The kinematical corrections in DVCS can be considered as a generalization of these target mass corrections.
However, higher-twist operators which can be reduced to the total derivatives of the leading-twist ones will also contribute to the DVCS reaction, since nonforward matrix elements are used.
As pointed out in Refs.\,\cite{Braun:2011dg,Braun:2011zr,Braun:2011th},
the distinction between two types of kinematical corrections is not Lorentz invariant and has no physical meaning. Both contributions should therefore better always be added together. Since the same operator governs the physics of the reaction \eqref{process}, one may use the same techniques to improve our understanding of the $s$-dependence of its amplitude. 
We thus study here the kinematical higher-twist corrections to the amplitude of the reaction \eqref{process},
in the kinematical domain suitable for a collinear QCD factorization framework where the leading-twist amplitude can be written as the convolution of a perturbatively calculable coefficient function and GDAs \cite{Diehl:2000uv}.

In Sect. II, we describe the kinematics of the $\gamma^* \gamma \to M \bar{M}$ process and recall the basic properties of GDAs. In Sect. III, we recall the results of Refs.\,\cite{Braun:2011dg,Braun:2011zr,Braun:2011th} and the definitions of the higher-twist kinematical operators. In Sect. IV, we derive the helicity amplitudes for the reaction (\ref{process}), including the kinematical higher-twist contributions. Sect. V shows our numerical estimates of the kinematical higher-twist contributions to the cross section for both $\pi \pi$ and $\eta \eta$ cases. We briefly present our conclusions in Sect. VI. Appendices A and B provide technical details for the calculation of helicity amplitudes.

\section{Kinematics and generalized distribution amplitudes }

\begin{figure}[htp]
\centering
\includegraphics[width=0.6\textwidth]{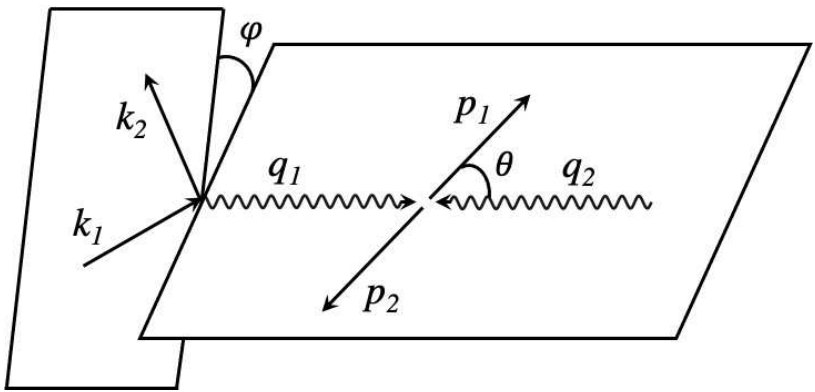}
\caption{Kinematics of the process $\gamma^*(q_1) \gamma(q_2) \to M(p_1) \bar{M}(p_2)$ in the center of mass of the meson pair; the virtual photon is emitted by the electron with four-momentum $q_1=k_1-k_2$.  }
\label{fig:angle}
\end{figure}

To describe the process \eqref{process}, we define the lightlike vectors $n$ and $\tilde{n}$  in a convenient way so that  they can be expressed by the momenta of the spacelike virtual photon $q_1$ and the  real photon $q_2$,
\begin{align}
\tilde{n} = q_1+(1-\tau)q_2, \qquad  n=q_2,
\label{eqn:lightlike}
\end{align}
where  $\tau=s/(Q^2+s)$, $Q^2=-q_1^2$, and $s=(q_1+q_2)^2=(p_1+p_2)^2$. The polar angle of the meson ($M$)  momenta $\theta$ is illustrated in Fig.\,\ref{fig:angle}, and is defined as
\begin{align}
\cos{\theta}=\frac{2q_1\cdot(p_2-p_1)}{\beta_0\,(Q^2+s)}, \qquad \beta_0=\sqrt{1-\frac{4 m^2}{s}}, 
\label{eqn:polar}
\end{align}
where $m$ is the meson mass.
For convenience, a new variable $\zeta_0=-\beta_0 \cos{\theta}$ is introduced
instead of $\cos\theta$
\begin{align}
\zeta_0= \frac{(p_2-p_1)\cdot n}{(p_2+p_1)\cdot n},
\label{eqn:skewness}
\end{align}
but the final amplitudes will be expressed in terms of $\cos{\theta}$. If the $z$-axis is chosen so that the $t$-$z$ plane contains the lightlike vectors $n$ and $\tilde{n}$, then only $\Delta=p_2-p_1$ has a transverse momentum, $\Delta=\zeta_0(\tilde{n}-\tau n) + \Delta_T$. Using the on-shell condition, we obtain $\Delta_T^2=4m^2-(1-\zeta_0^2)s$.

The amplitude for $\gamma^* \gamma \to M \bar{M} $ is defined as 
\begin{align}
A_{\mu \nu}=i\int d^4x\, e^{-ir\cdot x} \langle \bar{M}(p_2) M(p_1)  | \, 
T \{ j_{\mu}^{\text{em}}(z_1x)  j_\nu^{\text{em}} (z_2x) \} \, | 0 \rangle , \! \!
\label{eqn:amp0}
\end{align}
where $r=z_1q_1+z_2 q_2$, and the constraint $z_1-z_2=1$ is imposed for real constants $z_1$ and $z_2$. Owing to the electromagnetic gauge invariance, one can decompose this amplitude as \cite{Braun:2012bg}
\begin{align}
A^{\mu \nu}=-A^{(0)}\,g_{\perp}^{\mu\nu}+A^{(1)}\, \frac{\Delta_{\alpha}g_{\perp}^{\alpha \nu}}{Q}
\left(\tilde n^\mu+(1-\tau)n^\mu\right)
+\frac{1}{2}\, A^{(2)}\, \Delta_{\alpha}\Delta_{\beta}(g_{\perp}^{\alpha \mu} g_{\perp}^{\beta \nu}- \epsilon_{\perp}^{\alpha \mu}   \epsilon_{\perp}^{\beta \nu} )+ A^{(3) \mu}\, n^{\nu}
\label{eqn:amp1}
\end{align}
 with $ g_{\perp}^{\mu \nu}$ and $\epsilon_{\perp}^{ \mu \nu}$  given by
\begin{align}
g_{\perp}^{\mu \nu} =g^{\mu\nu}-
\frac{n^{\mu}\tilde{n}^{\nu}+n^{\nu}\tilde{n}^{\mu}}{n\cdot \tilde{n}}, \qquad    
\epsilon_{\perp}^{\mu \nu} = \epsilon^{\mu \nu \alpha \beta}\,
 \frac{\tilde{n}_{\alpha} n_{\beta}}{n\cdot \tilde{n}}.
\label{eqn:gt}
\end{align}
The last term in Eq.\,(\ref{eqn:amp1}) is of no interest since it does not contribute
to any observable, and the rest of them can be expressed in terms of the GDAs 
if the factorization conditions  $Q^2 \gg s, \Lambda_{\text{QCD}}^2$ are satisfied. The leading-twist amplitude was first presented in
Ref.\,\cite{Diehl:2000uv}
with the help of a twist-2 GDA $\Phi_q(z,\zeta_0, s)$ for an isoscalar meson pair,
\begin{align}
\langle \bar{M}(p_2) M(p_1)  | \,\bar{q}(z_1 n) \slashed{n}q(z_2 n)\, | 0 \rangle
=  2 P\cdot n \int dz \,e^{2i \left[ zz_1+(1-z)z_2 \right] P\cdot n}\, \Phi_q(z,\zeta_0, s),
\label{eqn:gda}
\end{align}
where $P=(p_1+ p_2)/2$, $\Phi_q$ is the  GDA for the quark flavor $q$, $\bar{q}(z_1 n) \slashed{n}q(z_2 n)$ is the leading-twist vector operator (a light-like Wilson line joining the points $z_1n$ and $z_2n$ is implied), and $z_1-z_2=1$ is not a necessary condition. This matrix element can alternatively be expressed in terms of double distributions (DDs) as 
\cite{Teryaev:2001qm}
\begin{align}
\langle \bar{M}(p_2) M(p_1)  | \, \bar{q}(z_1 n) \slashed{n}q(z_2 n)  \,  | 0 \rangle
=  \int d\beta\, d\alpha   \left[f_q(\beta, \alpha) \,\Delta\cdot n- g_q(\beta, \alpha)\, 2P\cdot n   \right] 
e^{-i  l_{z_1 z_2}\cdot n}
\label{eqn:dds}
\end{align}
with $f_q$ and $g_q$ having support on the rhombus $|\alpha|+|\beta|\leq 1$ and assumed to vanish at the boundary, and
\begin{align}
l_{z_1 z_2}=(z_2-z_1) \left[ \beta\, \frac{\Delta}{2} -(\alpha+1) P  \right] -2z_1 P.
\label{eqn:lz1z2}
\end{align}
 Then, one can easily relate the GDA to double distributions  
\begin{align}
\Phi_q(z,\zeta_0, s)=2\int d\beta \, d\alpha\,  \delta(y+\alpha-\beta\zeta_0 ) 
\left[f_q(\beta, \alpha)\, \zeta_0 - g_q(\beta, \alpha) \right], 
\label{eqn:dd-gda}
\end{align}
where $y=2z-1$. Since the meson pair is produced with charge conjugation $\mathcal{C}=+1$, one can obtain the relations
\begin{align}
f_q(\beta, \alpha)&=f_q(\beta, -\alpha), \qquad \phantom{--}g_q(\beta, \alpha)=-g_q(\beta, -\alpha), \\ \nonumber
f_q(\beta, \alpha)&=-f_q(-\beta, -\alpha),\qquad g_q(\beta, \alpha)=-g_q(-\beta, -\alpha)
\label{eqn:char-sy}
\end{align}
from charge conjugation invariance. Assuming that the DDs vanish at the boundaries, Eq.\,(\ref{eqn:dds}) can be put in the form
\begin{equation}
\langle \bar{M}(p_2) M(p_1)  | \, \bar{q}(z_1 n) \slashed{n}q(z_2 n)  \, | 0 \rangle
=  \frac{2i}{z_{12}}  \int d\beta  \,d\alpha\, 
 \phi_q(\beta, \alpha) \,
 e^{-i l_{z_1 z_2}\cdot n}, \label{eqn:dds-mod}
\end{equation}
where the notation $z_{12}=z_1- z_2$ is used.
A new distribution
\begin{align}
\phi_q(\beta, \alpha) =  \partial_{\beta}f_q(\beta, \alpha) + \partial_{\alpha}g_q(\beta, \alpha)   
\label{eqn:dds-new}
\end{align}
is introduced with symmetry $\phi_q(\beta, \alpha) = \phi_q(\beta,-\alpha) =\phi_q(-\beta, -\alpha)$, in order to simplify the calculation of the amplitudes thanks to the property
\begin{align}
\int d\beta\,d\alpha\,  \phi_q(\beta, \alpha)  \left[ a +b \,\alpha^n +c \,\beta^m   \right] =0, 
\label{eqn:dds-sys}
\end{align}
where $a$, $b$ and $c$ are constants which are independent of $\alpha$ and 
$\beta$, and the exponents $n$ and $m$ are odd numbers. Although the intermediate calculations involve the DD $\phi_q(\alpha,\beta)$, the final results will be presented in terms of the GDA using
\begin{align}
\frac{\partial \Phi_q(z,\zeta_0, s)}{\partial z} =4\int d\beta\, d\alpha\, \delta((2z-1)+\alpha-\beta \zeta_0)\, \phi_q(\beta, \alpha).
\label{eqn:dds-gda}
\end{align}

\section{Operator product expansion and helicity amplitudes }

A separation of kinematical and dynamical contributions in the time-ordered product of two electromagnetic currents
$i\,T \{j_{\mu}^{\text{em}}(z_1x)j_{\nu}^{\text{em}} (z_2x) \}$  was recently proved in Refs.\,\cite{Braun:2011dg,Braun:2011zr,Braun:2011th}.
The kinematical contributions only involve  the leading-twist distributions, 
whereas unrelated genuine higher-twist distributions are necessary for the dynamical contributions. One can thus improve the description of  reactions where two photons are involved by including the kinematical corrections, without any knowledge of the higher-twist distributions. 
A complete calculation of 
kinematical corrections was performed up to the twist-4 accuracy  for DVCS with a (pseudo)scalar target in  Ref.\,\cite{Braun:2012bg}. In this work we shall apply similar techniques to calculate the  kinematical 
higher-twist contributions in the $s$-$t$ crossed channel of DVCS, namely the reaction  $\gamma^* \gamma \to M \bar{M}$. The kinematical contributions to the operator $i\,T\{j_{\mu}^{\text{em}}(z_1x)j_{\nu}^{\text{em}} (z_2x) \}$ were given to twist-4 accuracy by \cite{Braun:2012bg,Braun:2011dg,Braun:2011zr},
\begin{align}
T_{\mu \nu}=\frac{-1}{\pi^2 x^4 z_{12}^3}
\left \{
x^{\alpha } \left[S_{\mu \alpha \nu \beta   } 
\mathbb{V}^{\beta}-i\epsilon_{\mu \alpha \nu \beta }\mathbb{W}^{\beta}
\right]  
+x^2 \left[ (x_{\mu} \partial_{\nu} + x_{\nu} \partial_{\mu}   )  \mathbb{X} + (x_{\mu} \partial_{\nu} - x_{\nu} \partial_{\mu}   )  \mathbb{Y} \right]
 \right \},
\label{eqn:kine-t4}
\end{align}
where the convention $\epsilon_{0123}=1$  is adopted for the antisymmetric tensor 
and $S^{ \mu \alpha \nu \beta }$ is defined as 
\begin{align}
S^{ \mu \alpha \nu \beta }=g^{\mu \alpha} g^{\nu \beta}-g^{\mu \nu}g^{\alpha \beta}+g^{\mu \beta} g^{\nu \alpha}.
\label{eqn:app2s}
\end{align}
In Eq.\,(\ref{eqn:kine-t4}), $\mathbb{V}_{\mu}$  and $\mathbb{W}_{\mu}$  contain contributions of twist 2, twist 3 and twist 4, whereas 
$\mathbb{X}$ and $\mathbb{Y}$ are purely twist 4, see Appendix \ref{app-a} for the detailed expressions.
In practice, the spinor formalism \cite{Braun:2008ia, Braun:2009vc} is used to calculate the amplitudes, since the expression of $T_{\mu \nu}$ becomes more compact and it is easier to figure out the twist of each term in the corresponding matrix elements.

In order to calculate the helicity amplitudes
of Eq.\,(\ref{eqn:amp1}), the photon polarization vectors are required. 
Choosing the momentum of the virtual photon along the z-axis, as shown in Fig.\,\ref{fig:angle},
its polarization vectors read \cite{Diehl:2000uv}
\begin{align}
&\epsilon_{0}^{\mu}=\frac{1}{Q}(|q_1^3|, 0, 0, q_1^0), \quad
\epsilon_{\pm}^{\mu}=\frac{1}{\sqrt{2}}(0, \mp 1, -i, 0),
\label{eqn:pol-virt}
\end{align}
where the lower indices $\pm$ and $0$ indicate the helicities of the photon.
The polarization vectors $\tilde{\epsilon}$ of the real photon only have the transverse components, and they are 
related to the ones of   the virtual photon as $\tilde{\epsilon}_{\pm}=- \epsilon_{\mp}$. 
In the reaction  $\gamma^* \gamma \to M\bar M $, the helicity amplitudes are defined as
\begin{align}
A_{i j}=  \epsilon_{i}^{\mu} \tilde{\epsilon}_{j}^{\nu } A_{\mu \nu},
\label{eqn:am-hel}
\end{align}
and there are only three independent helicity amplitudes owing to parity invariance,
as one can check from Eq.\,(\ref{eqn:amp1}).
Here we choose the independent helicity amplitudes as $A_{++}$, $A_{0+}$ and $A_{-+}$, then one obtains
\begin{align}
A_{++}=A_{--}=A^{(0)}, \quad
A_{0+}=-A^{(1)} (\Delta \cdot \epsilon_{-}), \quad
A_{-+}=-A^{(2)} (\Delta \cdot \epsilon_{-})^2.
\label{eqn:am-hel-ind}
\end{align}

At leading twist, the operator product expansion of $i\,T\{j_{\mu}^{\text{em}}(z_1x)j_{\nu}^{\text{em}} (z_2x) \}$ leads to the nonlocal operator
\begin{align}
O_{++}(z_1 n, z_2 n)= \sum_q e_q^2\,\bar{q}(z_1 n) \slashed{n} q(z_2 n)
\label{eqn:optb2}
\end{align}
with a lightlike separation.
Since we are interested in the reaction $\gamma^{\ast} \gamma \rightarrow M \bar{M} $ with a charge conjugation even final  state, one can safely neglect the contribution of the strange quark in the case of a $\pi$ meson pair,
\begin{align}
O_{++}(z_1 n, z_2 n)= e_u^2 \,\bar{u}(z_1 n) \slashed{n}u(z_2 n)+ e_d^2\, \bar{d}(z_1 n) \slashed{n}d(z_2 n) =\chi \left[ \bar{u}(z_1 n) \slashed{n}u(z_2 n)+ \bar{d}(z_1 n) \slashed{n}d(z_2 n) \right],
\label{eqn:opta2}
\end{align}
where $\chi= 5e^2/18$ is obtained thanks to the isospin symmetry\footnote{The amplitudes associated with $\bar{u}(z_1 n) \slashed{n}u(z_2 n)$ and $\bar{d}(z_1 n) \slashed{n}d(z_2 n)$ are the same for an isosinglet $\pi \pi$ state.}. One needs however to add $e_s^2\, \bar{s}(z_1 n) \slashed{n}s(z_2 n)$ to the operator
 $O_{++}(z_1 n, z_2 n)$
in the case of a $K$ meson pair.
The   kinematical  higher-twist  contributions in  the
operator product expansion of
$i\,T\{j_{\mu}^{\text{em}}(z_1x)j_{\nu}^{\text{em}} (z_2x) \}$ are related to the  operator 
$\mathcal{O}_{++}^{t=2}(z_1 , z_2 ) $,
where the separation $x$ is now not necessarily lightlike. We thus need to use the leading-twist projector
$\Pi(x, n)$ defined in Refs.\,\cite{Braun:2011dg,Braun:2011zr,Braun:2011th},
\begin{align}
\langle \bar{M}(p_2) M(p_1)  | \, \mathcal{O}_{++}^{t=2}(z_1 , z_2 )  \, | 0 \rangle
= \Pi(x, n) \langle \bar{M}(p_2) M(p_1)  | \, O_{++}(z_1 n, z_2 n)  \,  | 0 \rangle.
\label{eqn:projector}
\end{align}
Since the dependence on $n$ is always carried by a function of the type $e^{-il\cdot n}$ in Eq.\,(\ref{eqn:dds-mod}), the action of the leading-twist projector is simply given by
\begin{align}
\left[\Pi e^{-il\cdot n} \right](x)=e^{-il\cdot x}+\frac{x^2 l^2}{4} \int_0^1 dv\, v\,
e^{-ivl\cdot x}+\mathcal O(x^4).
\label{eqn:projector1}
\end{align}
Up to $1/Q^2$-accuracy, one obtains
\begin{align}
\langle \bar{M}(p_2) M(p_1)  | \, \mathcal{O}_{++}^{t=2}(z_1 , z_2 )  \,  | 0 \rangle
= \chi \, \frac{2i}{z_{12}}  \int d\beta\,  d\alpha 
\,\phi(\beta, \alpha) \left[ e^{-i l_{z_1 z_2}\cdot x}   + \frac{x^2 l_{z_1z_2}^2}{4} \int_0^1  dv\, v  \,
e^{-i v l_{z_1 z_2}\cdot x}  \right ],
\label{eqn:dds-t4}
\end{align}
where $\phi=\phi_u+\phi_d$  and  the second term provides a twist-4 contribution.
In addition to the leading-twist operator $\mathcal{O}_{++}^{t=2}(z_1 x, z_2 x)$, there are also the higher-twist operators
\begin{align}
\mathcal{O}_1(z_1, z_2)&= \left[i \mathbf{P}^{\mu}, \, \left[i \mathbf{P}_{\mu}, \, \mathcal{O}_{++}^{t=2}(z_1 , z_2 )\right] \right], \nonumber \\
\mathcal{O}_2(z_1, z_2)&= 
\left[i \mathbf{P}^{\mu}, \, \frac{\partial}{\partial x^{\mu}}\mathcal{O}_{++}^{t=2}(z_1 , z_2 ) \right], 
\label{eqn:total-deri}
\end{align}
which contribute to kinematical higher-twist corrections.
Using Eq.\,(\ref{eqn:dds-t4}), the matrix elements of   
$\mathcal{O}_1$ and $\mathcal{O}_2$ can be expressed up to $1/Q^2$-accuracy as 
\begin{align}
\langle \bar{M}(p_2) M(p_1)  | \, \mathcal{O}_1(z_1, z_2) \,  | 0 \rangle
&= -\chi \, \frac{2i}{z_{12}}\, s  \int d\beta\,d\alpha   \,
\phi(\beta, \alpha) \, e^{-i l_{z_1 z_2}\cdot x}, \nonumber \\
\langle \bar{M}(p_2) M(p_1)  | \, \mathcal{O}_2(z_1, z_2) \,  | 0 \rangle
&= \chi\,  \frac{2i}{z_{12}}  \int d\beta\,d\alpha   \,
\phi(\beta, \alpha) \left[ 2P\cdot l_{z_1 z_2} \,  e^{-i l_{z_1 z_2}\cdot x}   
+i P \cdot x \,l_{z_1z_2}^2 \int_0^1  dv\, v  \,
e^{-i v l_{z_1 z_2}\cdot x}  \right ].
\label{eqn:o1o2-t4}
\end{align}
Since the operators $\mathcal{O}_1$ and $\mathcal{O}_2$ contain total derivatives, their matrix elements vanish in the forward limit and need not be considered in DIS. They provide however corrections of order $m^2/Q^2$ and $s/Q^2$ in the reaction $\gamma^* \gamma \to M \bar{M}$.

\section{helicity amplitudes in terms of GDAs}
In the following we will calculate the helicity amplitudes of $\gamma^* \gamma \to M \bar{M} $, 
adopting similar techniques to the ones used for DVCS in Ref.\,\cite{Braun:2012bg}.
There are three independent helicity amplitudes, which can be expressed in terms of DDs, 
\begin{align}
A_{0+}&=2  \chi \,  \frac{ \Delta \cdot \epsilon_{-} }{Q}
\int d\beta \, d\alpha\, \phi(\beta, \alpha) \, \beta\,
\frac{\ln(F)  }{F-1},  \nonumber \\
A_{-+}
&= -\chi \, \frac{(\Delta \cdot \epsilon_{-})^2}{2 n \cdot\tilde{n}}  \int d\beta   \, d\alpha\, \phi(\beta, \alpha) \, \beta^2
\partial_F \! \left[ \frac{1-2F}{F-1} \ln(F) \right],   \nonumber \\
A_{++}&= \chi   \int d\beta\,   d\alpha\,  \phi(\beta, \alpha)
\left\{ 2 \ln(F)  +  \left[ \frac{s}{n \cdot\tilde{n}}\,(F-\alpha) + \frac{\beta^2 \Delta_T^2  }{4n\cdot \tilde{n}} \,\partial_F    \right]   \frac{1}{F-1}   \left[ \frac{\ln(F)}{2}-\text{Li}_2(1)+  \text{Li}_2(F) \right]    \right\},
\label{eqn:amp-all}
\end{align}
where $\Delta_T^2=g_{\perp}^{\mu \nu} \Delta_{\mu}\Delta_{\nu}$ 
and $i \epsilon$ is omitted in the functions of $\ln$ and $\text{Li}_2$ since it will not contribute to the amplitudes.
Details of the calculations can be found in Appendix \ref{app-b}.
$A_{0+}$ and  $A_{-+}$ are proportional to $\Delta \cdot \epsilon_{-}$ and 
$(\Delta \cdot \epsilon_{-})^2$ as indicated by Eq.\,(\ref{eqn:am-hel-ind}), respectively,
and the amplitudes do not depend on $z_1$ and $z_2$ which indicates that the translation invariance is recovered  in the physical amplitudes. The function $F(\alpha, \beta)$ is defined as
\begin{align}
F(\alpha, \beta) = \frac{\alpha -\beta \zeta_0 +1}{2},
\label{eqn:F}
\end{align}
 where $F=0$ and $F=1$ 
correspond to the quark momentum fractions $z=1$ and $z=0$ of the GDAs, respectively.

We notice that there are three types of integrals  expressed by DDs in the obtained amplitudes, namely
\begin{align}
I_1&= \int d\beta\, d\alpha \,\phi(\beta, \alpha)\, Y(F), \nonumber \\
I_2&= \int d\beta\, d\alpha\, \phi(\beta, \alpha) \,\beta Y(F), \nonumber \\
I_3&= \int d\beta\, d\alpha\, \phi(\beta, \alpha)\,\beta^2  \partial_F Y(F),
\label{eqn:int-dds}
\end{align}
where $Y(F)$ is some function of $F$.  Inserting the identity 
$\int dy \,\delta(\beta \zeta_0 -y- \alpha)=1$ into the integrals above,  one can reexpress the integrals in terms of GDAs by using Eq.\,(\ref{eqn:dds-gda}), 
\begin{align}
I_1&= - \frac{1}{2} \int_0^1 dz \, \Phi(z,\zeta_0, s) \,\partial_z Y(1-z)  , \nonumber \\
I_2&= - \frac{\partial}{\partial \zeta_0} \int_0^1 dz  \,\Phi(z, \zeta_0, s)\,  Y(1-z), \nonumber \\
I_3&= 2 \, \frac{\partial^2}{\partial \zeta_0^2} 
\int_0^1 dz\, \Phi(z, \zeta_0, s) \,Y(1-z),
\label{eqn:int-gdas}
\end{align}
where $y=2z-1$ and $\zeta_0=-\beta_0 \cos{\theta}$ as defined in Eq.\,(\ref{eqn:polar}). Therefore, we can write the helicity amplitudes as
\begin{align}
A^{(0)}&= \chi \left\{  \left(1- \frac{s}{2Q^2}\right) \int_0^1 dz\, \frac{\Phi(z, \eta, s)}{1-z} 
-\frac{s}{Q^2}\int_0^1  dz \, \frac{\Phi(z,\eta, s)}{z}\, \ln(1-z)
\right. \nonumber \\
&\qquad -\left. \left(\frac{2s}{Q^2} \,\eta   +\frac{\Delta_T^2}{\beta_0^2 Q^2} \frac{\partial}{\partial \eta} \right)
\frac{\partial}{\partial \eta}
  \int_0^1 dz \,\frac{ \Phi(z,\eta, s) }{z} \left[ \frac{\ln(1-z)}{2} +\text{Li}_2(1-z) -\text{Li}_2(1)  \right]
  \right \}, \nonumber \\
A^{(1)}&=  \frac{2\chi}{\beta_0 Q} \frac{\partial}{\partial  \eta }   \int_0^1 dz \, \Phi(z, \eta, s) \,
\frac{\ln(1-z)}{z}, \nonumber \\
A^{(2)}&=-   \frac{2 \chi}{ \beta_0^2 Q^2} \frac{\partial^2}{\partial \eta^2}       \int_0^1 dz  \,\Phi(z, \eta, s) \,
\frac{2z-1}{z}\, \ln(1-z),
\label{eqn:int-gdas-a}
\end{align}
where $\eta=\cos \theta$ and $\Phi=\Phi_u+\Phi_d$. The GDA for $s$ quarks is also required in some reactions such as
$\gamma^{\ast} \gamma \rightarrow K^0 \bar K^0$ with a charge conjugation-even $K$ meson pair,
and we just need to replace $\chi \Phi$ with 
$e_u^2 \Phi_u+e_d^2 \Phi_d +e_s^2 \Phi_s$ in the above amplitudes.
One can clearly see the $\mathcal O(s/Q^2)$ corrections  in the amplitudes, and the target mass correction of order $\mathcal O(m^2/Q^2)$ is implicit since it appears
in the term $\Delta_T^2/ Q^2$ by considering $\Delta_T^2=4m^2-(1-\zeta_0^2)s$.
In general, charge conjugation-even GDAs can be expanded as \cite{Diehl:2000uv}
\begin{align}
\Phi(z, \cos \theta, s)=6\, z(1-z) \sum_{\substack{ n=1\\ n\,\text{odd} }}^{\infty} \sum_{\substack{l=0\\ l\,\text{even}}}^{n+1}
\tilde{B}_{nl}(s) \,C_n^{(3/2)}(2z-1)\, P_l(\cos \theta),
\label{eqn:gda-expression}
\end{align}
where $C_n^{(3/2)}(x)$ are Gegenbauer polynomials and 
$P_l(x)$ are Legendre polynomials.
Due to this general expression for GDAs, the singularities of $\frac{1}{z}$, 
$\frac{1}{1-z}$ and $\ln{(1-z)}$ in the helicity amplitudes will be compensated by the GDA when 
$ z\rightarrow 0$ and $ z\rightarrow 1$. As a consequence, the amplitudes have no end-point singularities.
In the  asymptotic limit ($Q^2 \rightarrow \infty$), only the terms with $n=1$  survive,
\begin{align}
\Phi(z, \cos \theta, s)=18\, z(1-z) (2z-1) \left[\tilde{B}_{10}(s)
+\tilde{B}_{12}(s)  P_2(\cos \theta) \right],
\label{eqn:gda-exp-asm}
\end{align}
where the first  and second terms correspond to the S-wave and
D-wave production of a meson pair, respectively. The nonvanishing helicity-flip
amplitudes $A_{-+}(A^{(2)})$ and $A_{0+}(A^{(1)})$ indicate the existence of a D-wave GDA.

\section{Numerical estimates of the higher-twist kinematical contributions}

\subsection{$\pi \pi$ GDA  extracted from Belle measurements  }
\begin{figure}[ht]
\centering
\includegraphics[width=0.85\textwidth]{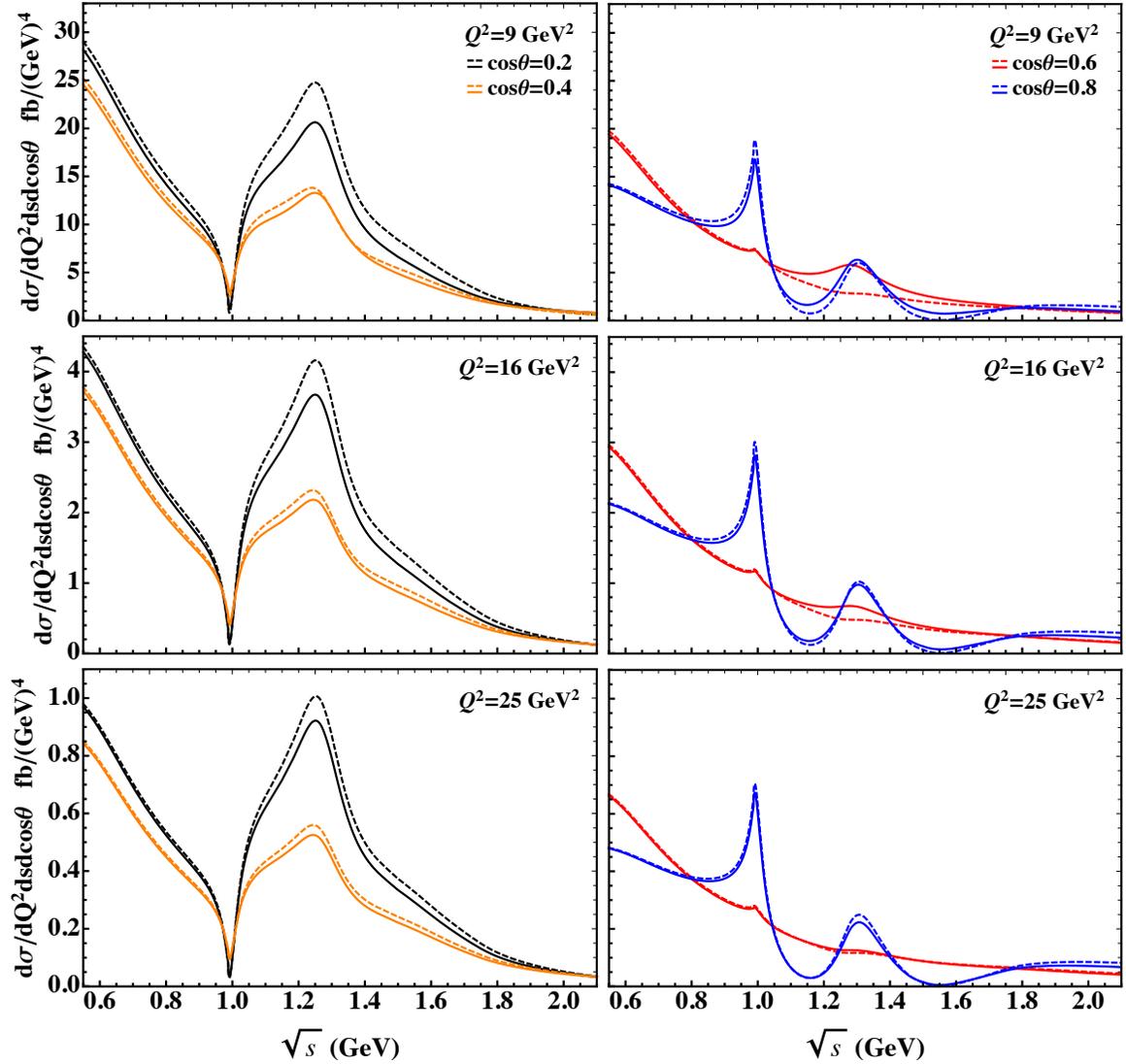}
\caption{ Differential cross section for  $e \gamma \to e \pi^0  \pi^0 $ calculated with   $\pi \pi$ GDA  extracted from Belle measurements through a leading-twist analysis \cite{Kumano:2017lhr}. Dashed curves show the twist-2 results, while solid curves include the kinematical higher-twist contributions. The selected values are $s_{e\gamma} = 30$ GeV$^2$, $Q^2 = 9~(16, 25)$ GeV$^2$ and $\cos \theta = 0.2~(0.4, 0.6, 0.8)$ as indicated on the different panels.}
\label{fig:num}
\end{figure}

\begin{figure}[htp]
\centering
\includegraphics[width=0.85\textwidth]{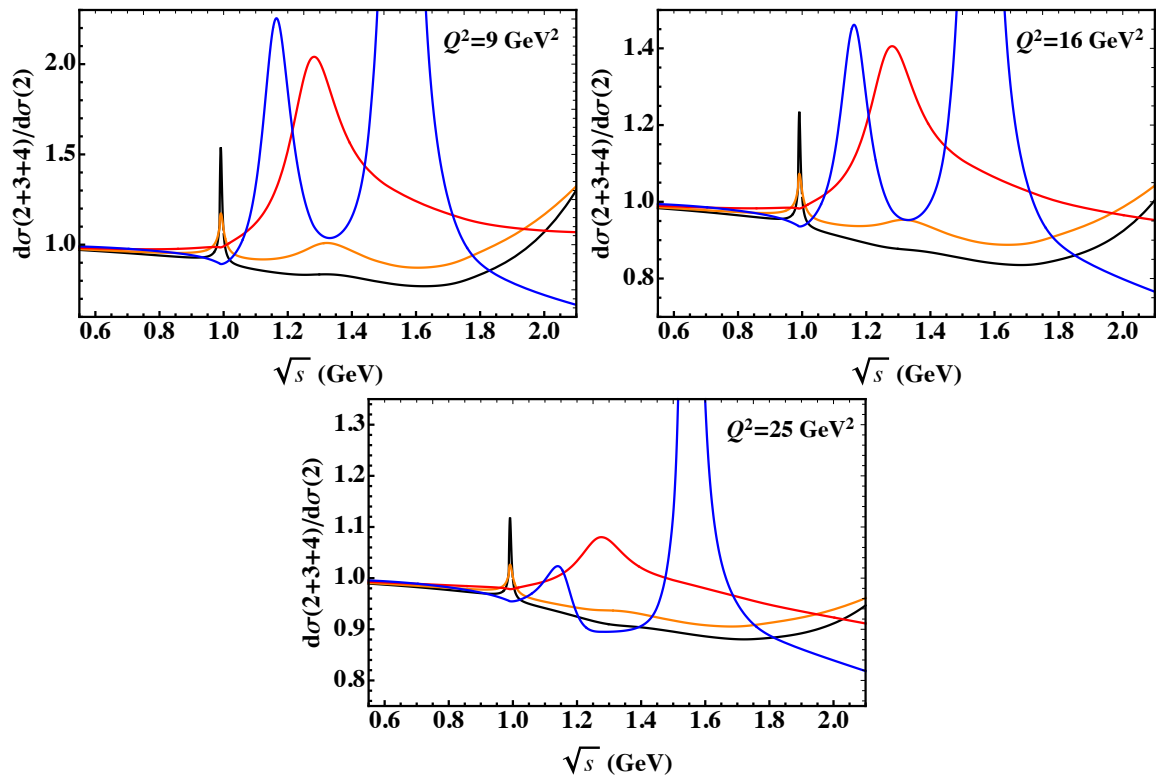}
\caption{ Ratio $d\sigma(2+3+4)/d\sigma(2)$ with the $\pi \pi$ GDA  extracted from Belle measurements, same conventions as in Fig.\,\ref{fig:num}.}
\label{fig:num2}
\end{figure}

The process  $\gamma^* \gamma \to M \bar{M} $ can be measured in  $e^+ e^-$ collisions, which are accessible at KEKB and SuperKEKB. In Ref.\,\cite{Diehl:2000uv}, the differential cross section for $e \gamma \to e M \bar{M}$ is expressed as 
\begin{align}
\frac{d \sigma}{dQ^2 \,ds\, d(\cos \theta)\, d\varphi}&=
\frac{\alpha_{\text{em}}^3 \beta_0}{16 \pi s_{e\gamma}^2 } \,\frac{1}{Q^2(1-\epsilon)}   \,\Big[ |A_{++}|^2+ |A_{-+}|^2+2\epsilon\, |A_{0+}|^2 -2\sqrt{\epsilon(1+\epsilon)} \cos \varphi \, \text{Re}(A_{++}^{\ast} A_{0+} -A_{-+}^{\ast} A_{0+})  \nonumber \\ 
 &\quad   - 2 \epsilon \cos (2\varphi)\,  \text{Re}(A_{++}^{\ast} A_{-+})   \Big],
\label{eqn:epho-cro}
\end{align}
where $\varphi $ is the azimuthal angle of the meson pair as illustrated in Fig.\,\ref{fig:angle}
and $s_{e\gamma}$ is the center-of-mass squared energy of  
$e \gamma$. $\epsilon$ is defined as usual by 
\begin{align}
\epsilon=\frac{1-y}{1-y+y^2/2}\quad \text{with}\quad  y=\frac{Q^2+s}{s_{e \gamma}}.
\label{eqn:eplson}
\end{align}

In the reaction $e \gamma \rightarrow e \pi \pi$, there are two types of contributions to the cross section. The final $\pi^+ \pi^-$ with negative charge conjugation couples to a virtual photon, and its contribution is expressed in terms of the pion electromagnetic form factor. When the charge conjugation of $\pi \pi$ is positive, the pion pair can be $\pi^0 \pi^0$ or $\pi^+ \pi^-$. Using factorization, 
this type of contribution is determined by GDAs, which we are interested in.
The $\pi^+ \pi^-$ GDA is equal to the one of $\pi^0 \pi^0$ due to the isospin symmetry.  However, since 
$\pi^0 \pi^0$ are identical bosons,   $\cos \theta$  will be restricted to $0  \le \cos \theta \le 1$ in Eq.\,(\ref{eqn:epho-cro}). After integration over $\theta$ and $\varphi$, the cross section for $C$-even $\pi^+ \pi^-$ production is then twice as large as the one for $\pi^0 \pi^0$  .

 In 2016, the Belle Collaboration released the measurements of differential cross section for $\gamma^* + \gamma \rightarrow \pi^0+ \pi^0$ \cite{Belle:2015oin}.  Since the final state is $\pi^0 \pi^0$, there is no contribution from the pion electromagnetic form factor.
The twist-2 $\pi \pi$ GDA  was extracted  by using the leading-twist amplitude \cite{Kumano:2017lhr}.
  We use this  pion GDA to estimate the cross section for  $e \gamma \to e \pi  \pi $ where the integral over $\varphi$  is performed in Eq.\,(\ref{eqn:epho-cro}),
\begin{align}
\frac{d \sigma}{dQ^2\, ds\, d(\cos \theta) }=
\frac{\alpha_{\text{em}}^3 \beta_0}{8 s_{e\gamma}^2 } \frac{1}{Q^2(1-\epsilon)}   \left[ |A_{++}|^2+ |A_{-+}|^2+2\epsilon\, |A_{0+}|^2    \right].
\label{eqn:epho-cro1}
\end{align}
In order to show the size of the higher-twist kinematical contributions,  Eqs.\,(\ref{eqn:int-gdas-a}) and (\ref{eqn:epho-cro1})
are used to calculate the cross section, and the results are depicted as the solid lines in Fig.\,\ref{fig:num}.
The dashed lines represent the leading-twist cross sections.
Considering  the kinematics of Belle measurements, we choose the values  $Q^2=9, 16, 25$ GeV$^2$, $s \in (0.25, 4)$ GeV$^2$, and we set $s_{e \gamma}=30 $ GeV$^2$ which is the typical value at Belle. 
In Fig.\,\ref{fig:num}, black lines  denote $\cos \theta=0.2$ and orange lines correspond to $\cos \theta=0.4$, while $\cos \theta=0.6$ and  $\cos \theta=0.8$ are depicted as red and blue, respectively.
As $Q^2$ increases,  kinematical contributions become less important,
which is consistent with the fact that the kinematical contributions  are suppressed by $1/Q$ or $1/Q^2$.
The kinematical contributions cannot be neglected in the region where 
$ \sqrt{s} \geq 1$ GeV.
The helicity-flip amplitudes 
$A_{-+}$ and $A_{0+}$ receive only contributions from the D-wave GDA,  and a large difference between two types of cross sections is displayed around the D-wave resonance region of $f_2(1270)$ in Fig.\,\ref{fig:num}.
Hence, the study of the amplitudes $A_{-+}$ and $A_{0+}$ will be important for the investigation of this resonance region. 
The  kinematical higher-twist corrections contribute $\sim 15\%$ to the cross section on average, if one  restricts the process $e \gamma \to e \pi \pi$ to the kinematics  of Belle measurements.

In Fig.\,\ref{fig:num2}, we also present the ratio $d\sigma(2+3+4)/d\sigma(2)$
where $d\sigma(i)$ ($i=2,3,4$) is the  twist-$i$ contribution to the cross section,
and  the colors of the lines indicate different values of $\cos \theta$ as in Fig.\,\ref{fig:num}. In this figure, the contributions of  the kinematical higher-twist corrections are quite clear,
so we can infer that the kinematical corrections cannot be neglected when $\sqrt{s} > 1$ GeV.
Around $\sqrt{s} \sim 1.5$ GeV, the kinematical corrections are dominant in the cross section with $\cos \theta=0.8$; this appears because  the twist-2 cross section is quite tiny when calculated with the  GDA extracted from Belle measurements;  this GDA may however not be accurate in this region since the uncertainties of Belle measurements are quite large there;
this ratio may thus not reflect the real physics around $\sqrt{s} \sim 1.5$ GeV.

As we have seen, the kinematical corrections are not negligible in the region $\sqrt{s} > 1$ GeV, which turns out to be important for the studies of the pion EMT form factors. Indeed, since pion GPDs cannot easily be measured in experiments, GDAs offer a way to investigate the timelike EMT form factors of pions.
The spacelike EMT form factors can then be obtained from the timelike ones by using dispersion relations, in which case the timelike EMT form factors of $\sqrt{s} > 1$ GeV are needed to be included numerically. As a consequence, it is important to use the most accurate description of the cross section with the inclusion of kinematical contributions. 

As pointed out above, the uncertainties of Belle measurements \cite{Belle:2015oin} are quite large, and the statistical errors are dominant.
However, this situation will be improved substantially soon, since the Belle II collaboration just started taking data at the SuperKEKB with a much higher luminosity. Precise measurements of $\gamma^* + \gamma \rightarrow M+ \bar{M}$ are expected in the near future, and an accurate description of the amplitudes for the study of GDAs requires the inclusion of kinematical contributions up to twist 4.

\subsection{Asymptotic pion GDA}

\begin{figure}[htp]
\centering
\includegraphics[width=0.85\textwidth]{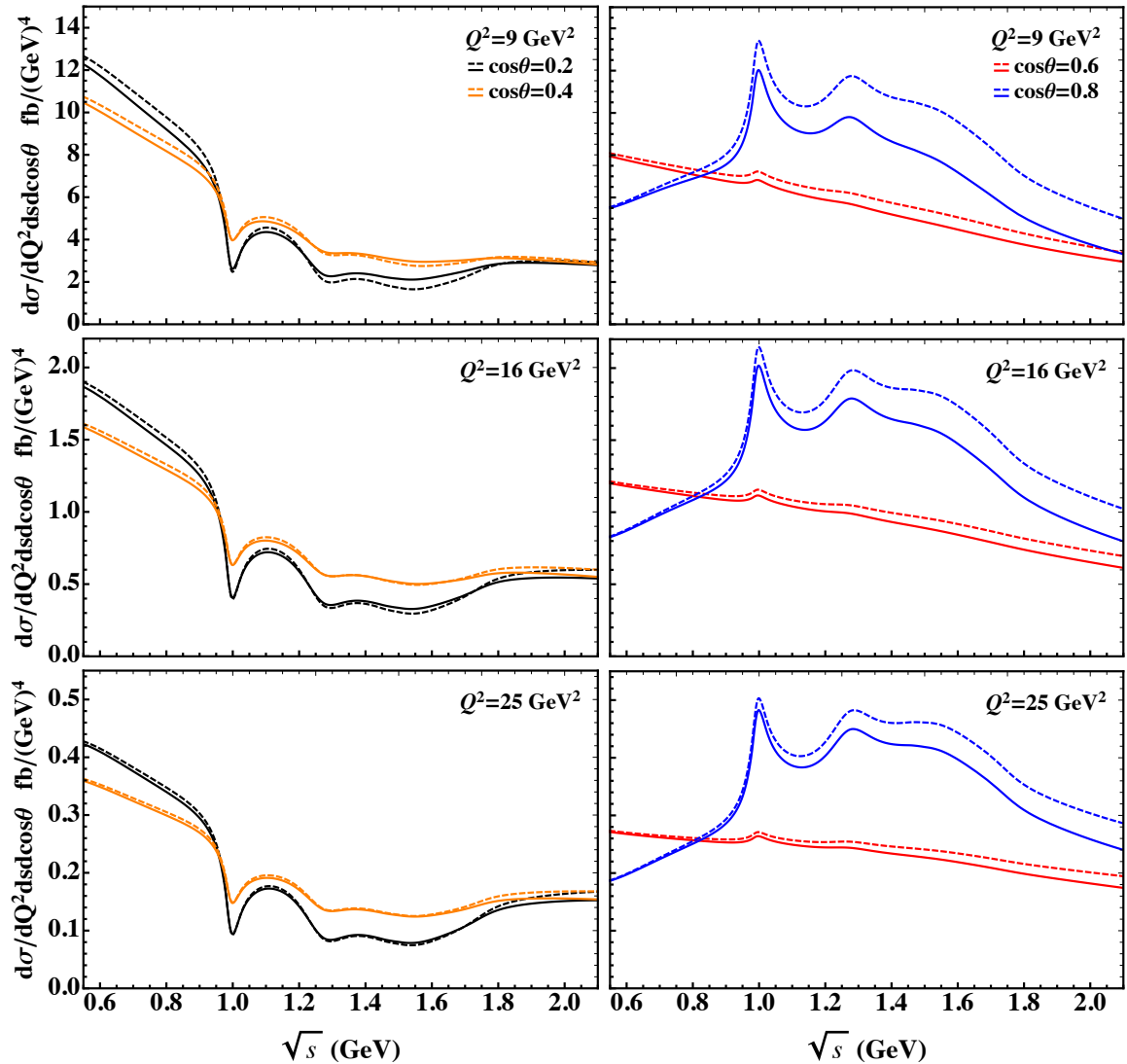}
\caption{ Differential cross section for  $e \gamma \to e \pi^0  \pi^0 $ with the asymptotic $\pi \pi$ GDA described in the text, same conventions as in Fig.\,\ref{fig:num}.}
\label{fig:num3}
\end{figure}

 \begin{figure}[htp]
\centering
\includegraphics[width=0.85\textwidth]{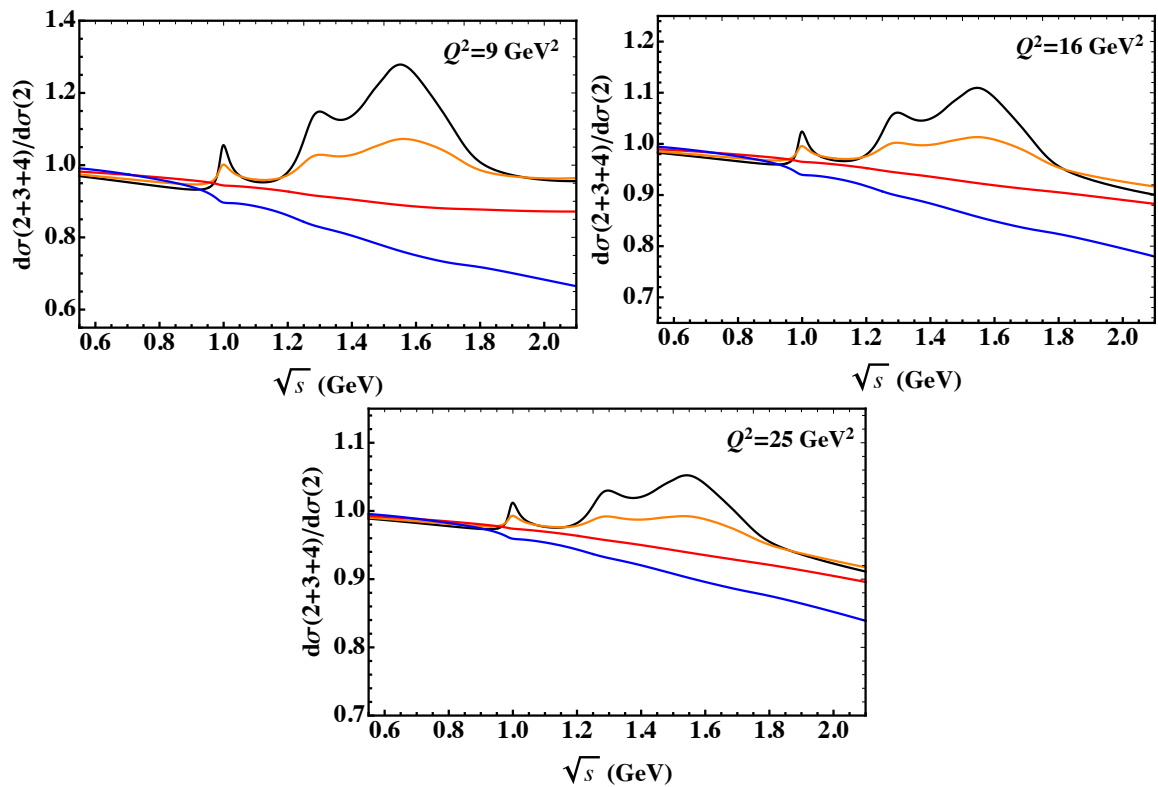}
\caption{ Ratio   $d\sigma(2+3+4)/d\sigma(2)$ with the asymptotic $\pi^0 \pi^0$ GDA described in the text, same conventions as in Fig.\,\ref{fig:num}.}
\label{fig:num4}
\end{figure}

The asymptotic pion GDA used in our calculation is taken from Eq.\,(68) of Ref.\,\cite{Diehl:1998dk},
\begin{align}
\Phi(z, \cos \theta, s)=20\, z(1-z)(2z-1) R_{\pi} \left[\frac{-3+\beta_0^2}{2}\, e^{i \delta_0} +
\beta_0^2 e^{i \delta_2}  P_l(\cos \theta)\right],
\label{eqn:gda-asy}
\end{align}
where $\delta_0$ and $\delta_2$ are $\pi \pi$ elastic scattering phase shifts in the isospin 0 channel \cite{Bydzovsky:2016vdx, Bydzovsky:2014cda, Surovtsev:2010cjf}.
$ R_{\pi}=0.5$ represents the  momentum fraction carried by quarks in the pion meson.
In this asymptotic GDA, we do not include the contribution of the $f_2$ resonance. However, we believe it is reasonable to use this  GDA here, since our purpose  is not to predict the cross section for $e \gamma \to e \pi^0 \pi^0$ precisely, but to
estimate the magnitude  of  kinematical higher-twist  contributions and determine whether one can neglect them or not in the cross section.

In Fig.\,\ref{fig:num3}, we show the cross section for  $e \gamma \to e \pi^0 \pi^0$ with fixed $Q^2$ and $\cos \theta$, the dashed lines are  the twist-2 cross sections, while the solid ones indicate the cross sections with kinematical contributions included. The colors of the lines denote different values of $\cos \theta$ as indicated on the different panels of the figure. Similarly to the case of the extracted $\pi \pi$ GDA,
the kinematical corrections are important to describe the cross section in the region of $\sqrt{s} > 1$ GeV.
As $Q^2$ increases, the kinematical contributions become less important. Compared with Fig.\,\ref{fig:num}, the magnitude of the cross sections
are similar at different $Q^2$, even though the asymptotic GDA is very different from the extracted  GDA from Belle measurements.
We also present the ratio of $d\sigma(2+3+4)/d\sigma(2)$  in Fig.\,\ref{fig:num4}, where the colors of the lines indicate different values of $\cos \theta$ as  in Fig.\,\ref{fig:num3}. In this figure, we can see that the kinematical contributions can account up to about $40\%$ of the cross section, which is of course not negligible.
Compared with  Fig.\,\ref{fig:num2}, the magnitude of the ratios 
are  slightly smaller than the ones obtained from extracted  GDA from Belle measurements.

\subsection{Model  for $ \eta \eta$ GDA  }

\begin{figure}[htp]
\centering
\includegraphics[width=0.85\textwidth]{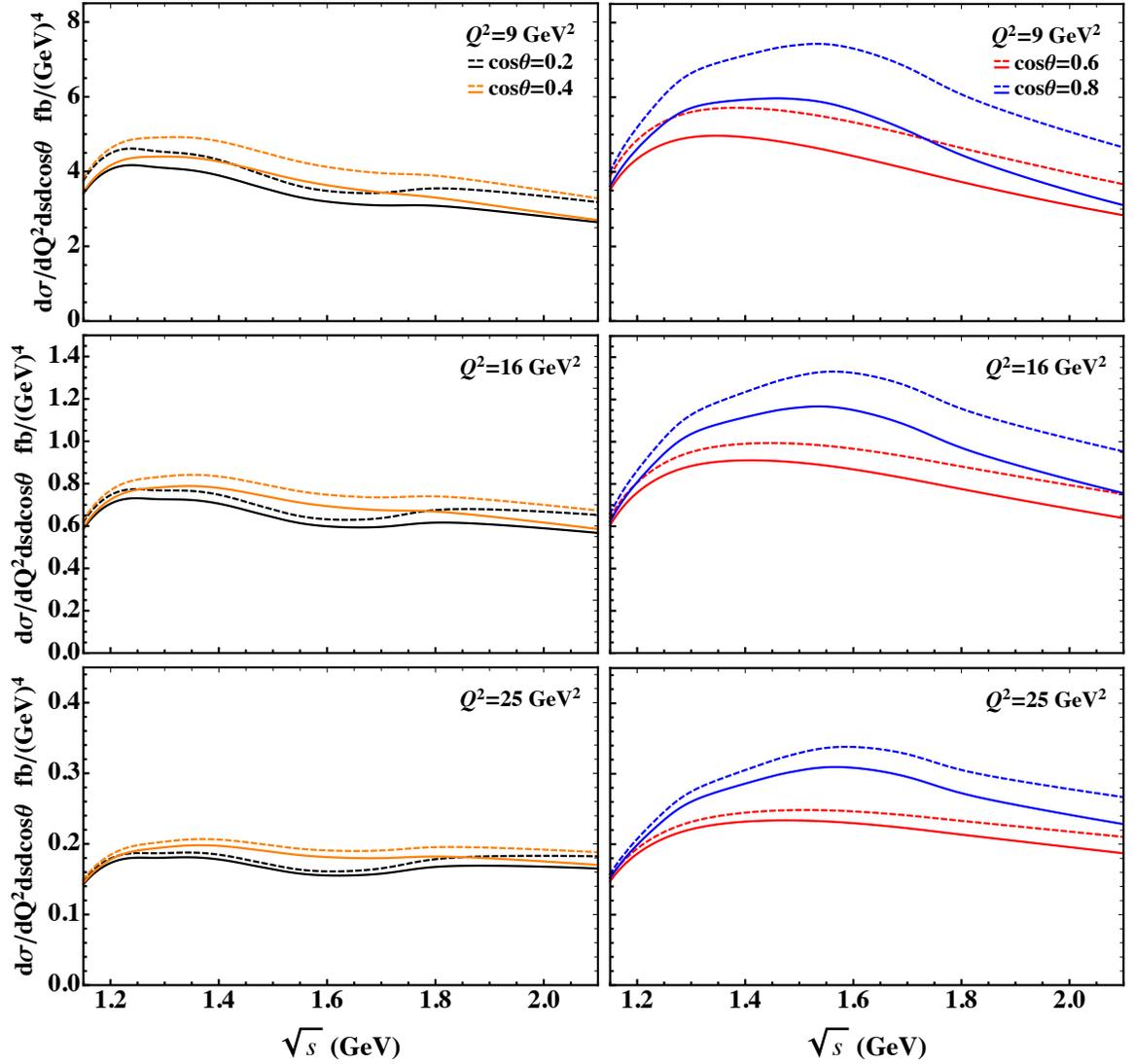}
\caption{ Differential cross section for  $e \gamma \to e \eta  \eta $ with the model $\eta \eta$ GDA described in the text, same conventions as in Fig.\,\ref{fig:num}.}
\label{fig:num5}
\end{figure}

\begin{figure}[htp]
\centering
\includegraphics[width=0.85\textwidth]{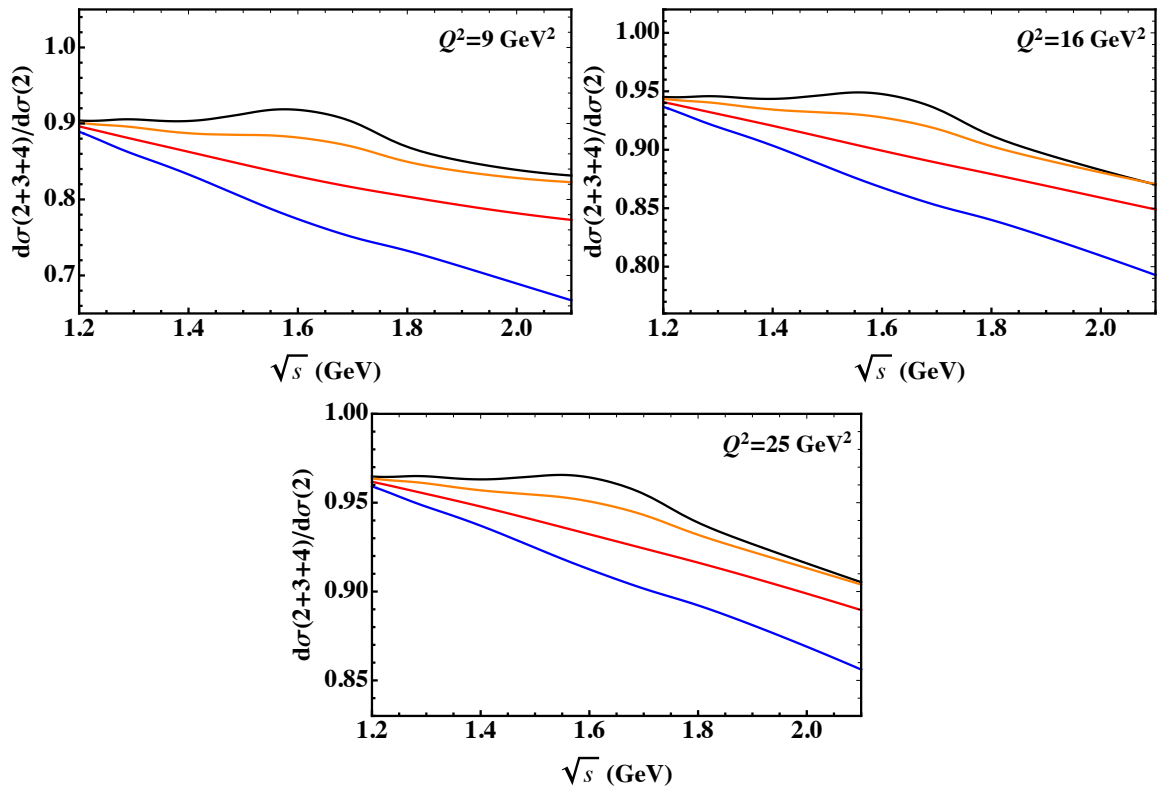}
\caption{Ratio $d\sigma(2+3+4)/d\sigma(2)$ with the model  $\eta \eta$ GDA described in the text, same conventions as in Fig.\,\ref{fig:num}.}
\label{fig:num6}
\end{figure}

As mentioned in the introduction, the kinematical contributions are expected to be proportional to $m^2/Q^2$ and $s/Q^2$. 
Since the mass of $\pi$ is quite small compared to the $Q^2$ values of Belle measurements, it is interesting to check the kinematical contributions  for the production of a pair of slightly heavier mesons, such as $K$ or $\eta$ mesons. 
On the one hand, this helps understand  how the target mass corrections of order $\mathcal O(m^2/Q^2)$ affect the cross section.
On the other hand, the $K \bar K$ and $\eta \eta$ GDAs  can also be measured by Belle and Belle II experiments;
the Belle Collaboration indeed released the cross section for  $\gamma^* + \gamma \rightarrow K_0+ \bar{K}_0$ in 2018 \cite{Belle:2017xsz}. It will therefore be necessary to investigate the kinematical corrections for the production of $K$ and $\eta$ meson pairs.
Unfortunately,  there is almost no information on these GDAs at present. 
Here, we just simply replace the mass of $\pi$ with the one of $\eta$ in Eq.\,(\ref{eqn:gda-asy}) and keep other parameters unchanged, then use this GDA to estimate the cross section and the ratio of various twist contributions  for  $e \gamma \to e \eta \eta$ with $s_{e \gamma}=30$ GeV$^2$.
The  differential cross section for $e \gamma \to e \eta \eta$ is shown in Fig.\,\ref{fig:num5}, and the ratio of  
$d\sigma(2+3+4)/d\sigma(2)$ is presented in Fig.\,\ref{fig:num6}.
The values of $Q^2$ are chosen as $Q^2=$ 9, 16 and 25 GeV$^2$ together with  1.2 GeV $  \leq  \sqrt{s} \leq $ 2.2 GeV, and the black (orange, red, blue) lines in 
Fig.\,\ref{fig:num5} and Fig.\,\ref{fig:num6}.
represent $\cos \theta=0.2$ (0.4, 0.6, 0.8).
The kinematical contributions account for somewhat less than $40\%$ of the cross section, which cannot be neglected either.
The kinematical higher-twist contributions have a significant impact on the cross section even in the region ($\sqrt{s} \sim 1.2$ GeV) which is close to the $\eta \eta$ threshold.
Compared with the ratios of Figs.\,(\ref{fig:num2}) and (\ref{fig:num4}), the  kinematical
higher-twist contributions are always negative, and those 
negative contributions can only come from the amplitude of $A_{++}$, since $A_{+-}$ and 
$A_{0-}$ always contribute to the cross section positively as indicated by Eqs.\,(\ref{eqn:int-gdas-a}) and (\ref{eqn:epho-cro1}).
The importance of the kinematical contributions does diminish as $m$ increases from the pion mass to the $\eta$ mass, simply because  the  negative 
kinematical higher-twist contributions from $A_{++}$ are compensated by the positive ones from $A_{0-}$ and $A_{+-}$.

\section{Summary}
In this paper we presented a complete calculation of kinematical higher-twist corrections for the helicity amplitudes of  the reaction $\gamma^* + \gamma \rightarrow M+ \bar{M}$ up to twist 4, where only the leading-twist GDA is involved in the description of the cross section. In case of $\pi \pi$ production, we use two types of GDAs to estimate the kinematical higer-twist contributions in the cross section, namely  the $\pi \pi$ GDA extracted from the Belle measurements and the asymptotic $\pi \pi$ GDA.
Even though those two GDAs are very different, both of them lead to kinematical corrections which cannot be neglected for $\gamma^* + \gamma \rightarrow  \pi \pi$ 
considering the kinematics accessible at Belle and Belle II.
Moreover, the relative magnitude of the higher-twist corrections is also comparable for the two types of $\pi \pi$ GDAs
at different values of $Q^2$, as seen from  
Figs.\,(\ref{fig:num2}) and (\ref{fig:num4}).

Due to the small pion mass, only  kinematical corrections of the type $\mathcal O(s/Q^2)$  contribute to the cross section for $\gamma^{\ast} + \gamma \rightarrow  \pi \pi$. 
The production of a pair of slightly heavier mesons is needed to check how the target mass corrections $m^2/Q^2$ affect the cross section.
Since the  $\eta \eta$ GDA is an unknown quantity at present, we calculate this effect with a model GDA 
identical to  the asymptotic $\pi \pi$ GDA except that the mass of $\eta$ is used. 
This calculation indicates that kinematical corrections are also not 
negligible in this case.
Furthermore, the negative kinematical corrections from the amplitude $A_{++}$ are dominant over the higher-twist contributions from
$A_{+-}$  and $A_{0-}$, and the kinematical corrections are always negative in the cross section, which is different from the $\pi \pi$ case where kinematical corrections can go both ways.

In conclusion, let us stress that while the uncertainties of present Belle measurements \cite{Belle:2015oin} are too large for our study to invalidate the conclusions of \cite{Kumano:2017lhr}, the situation should change substantially in a near future since Belle II collaboration just started taking data at the SuperKEKB with a much higher luminosity. An accurate description of the amplitudes in terms of GDAs will therefore require the inclusion of kinematical contributions up to twist 4. Note also that precise measurements of $\gamma^* + \gamma \rightarrow M+ \bar{M}$ for various mesons will be of utmost importance to address the questions of the pion EMT form factors and of  the impact-parameter representation of GDAs \cite{Pire:2002ut}.

In the future, this work can be extended to the production of 
other meson pairs, for example the $\gamma^{\ast} \gamma \rightarrow \pi \eta$ channel which should help unraveling the quark and gluon structure of hybrid meson ($J^{PC} = 1^{- +}$) \cite{Anikin:2006du}. The scattering amplitude is likely to be sensitive to sizeable kinematical higher twist contributions. This channel has recently been advocated \cite{Teryaev:2022pke} to be related to shear viscosity of quarks in hadronic matter. The production of a pair of vector mesons should also be discussed, opening the way to a meaningful extraction of the EMT form factors for $\rho$ or $\omega$ mesons.

Similar relations apply to the timelike process amplitude $\gamma^* \to  M \bar{M} \gamma $ which opens another access to GDAs \cite{Lu:2006ut} through the interference with the initial-state radiation amplitude in the process $e^+ e^- \to M \bar{M} \gamma$, as experimentally proven by the BABAR collaboration \cite{BaBar:2015onb}.
The extension of our work to the process $\gamma^* + \gamma \rightarrow N+ \bar N$ will also be needed if Belle II detector is able to detect this channel.

\section{Acknowledgements}
We acknowledge useful discussions with Lech Szymanowski, Pieter Taels, Oleg Teryaev,    Sadaharu Uehara and Jakub Wagner.
Qin-Tao Song is supported by the National Natural Science Foundation 
of China under Grant Number 12005191 and the China Scholarship Council 
for visiting Ecole Polytechnique.

\appendix
\section{Expressions for the T-product of two electromagnetic currents}
\label{app-a}
Here we give the detailed expressions of $\mathbb{V}_{\mu}$, $\mathbb{W}_{\mu}$, $\mathbb{X}$ and $\mathbb{Y}$ \cite{Braun:2011dg, Braun:2011th}, which are used in the T-product of two electromagnetic currents in Eq.\,(\ref{eqn:kine-t4}).
\begin{align}
\mathbb{V}_{\mu}(z_1, z_2)&=\mathfrak{B}_{\mu}(z_1, z_2)-\mathfrak{B}_{\mu}(z_2, z_1)+x_{\mu}  \Delta\mathbb{A}(z_1, z_2), \nonumber \\
\mathbb{W}_{\mu}(z_1, z_2)&=-\mathfrak{B}_{\mu}(z_1, z_2)-\mathfrak{B}_{\mu}(z_2, z_1),\nonumber \\
\mathbb{X}(z_1, z_2)&=\mathbb{C}(z_1, z_2)-\mathbb{C}(z_2, z_1), \nonumber \\
\mathbb{Y}(z_1, z_2)&=-\mathbb{C}(z_1, z_2)-\mathbb{C}(z_2, z_1),
\label{eqn:app1}
\end{align}
where $\Delta\mathbb{A}(z_1, z_2)=\mathbb{A}(z_1, z_2)-\mathbb{A}(z_2, z_1)$  is a pure twist-4 operator,
\begin{align}
\mathbb{A}(z_1, z_2)&=\frac{1}{4} 
\int_0^1  du
\left\{  u^2 \ln{u}\, z_1 z_2\, \mathcal{O}_{1}(z_1u,  z_2 u)
+\left[\left(z_2 \partial_{z_2}-\frac{z_1}{z_{12}} -\ln{u} \,z_2\partial_{z_2}^2z_{12}\right)\mathcal{R}(uz_1,uz_2)-(1\leftrightarrow 2)\right] \right\}.
\label{eqn:app5}
\end{align}
The function $\mathcal{R}(z_1,z_2)$ is related to  the total derivative operators $ \mathcal{O}_1$ and $ \mathcal{O}_2$ through
\begin{align}
 \mathcal{R}(z_1,z_2)=z_{12} \int_{z_2}^{z_1} \frac{dw}{z_{12}} \int_{z_2}^{w} \frac{dw_1}{z_{12}}
\, \frac{w_1-z_2}{z_1-w_1} \left[ \frac{1}{2}\,S_{+}\mathcal{O}_1(w, w_1)-(S_0-1)\mathcal{O}_2(w, w_1)
 \right],
 \label{eqn:ope-rz1z2}
\end{align}
where $S_{+}$ and $S_0$ are differential operators of $w$ and $w_1$,
\begin{align}
S_{+}&=w^2 \partial_w +2w +w_1^2 \partial_{w_1} +2w_1, \nonumber\\
S_{0}&=w \partial_w +w_1 \partial_{w_1} +2.
\label{eqn:ope-diff}
\end{align}
The operator $\mathfrak{B}_{\mu}(z_1, z_2)$ contains all twists starting from twist 2,
\begin{align}
\mathfrak{B}_{\mu}(z_1, z_2)=\mathfrak{B}_{\mu}^{t=2}(z_1, z_2)+\mathfrak{B}_{\mu}^{t=3}(z_1, z_2)+
\mathfrak{B}_{\mu}^{t=4}(z_1, z_2).
\label{eqn:appa}
\end{align}
The  twist-2 and twist-3 parts are defined as
\begin{align}
\mathfrak{B}_{\mu}^{t=2}(z_1, z_2)=&\frac{1}{2}\, \partial_{\mu} \int_0^1 du  \,\mathcal{O}_{++}^{t=2}(uz_1,  uz_2 ), \nonumber \\
\mathfrak{B}_{\mu}^{t=3}(z_1, z_2)=&\frac{1}{4} \int_0^1 du\,u\int_{z_2}^{z_1} \frac{dw}{z_{12}}
\left\{   \left[ i \mathbf{P}^{\nu},  K_{\alpha \nu \beta \mu }(x^{\alpha} \partial^{\beta})z_1 \mathcal{O}_{++}^{t=2}(z_1u,  w u) +
K_{ \mu \alpha \nu \beta } (x^{\alpha} \partial^{\beta})z_2 \mathcal{O}_{++}^{t=2}(wu,  z_2 u) \right] \right.  \nonumber \\
& \left.+  \ln(u)\, \partial_{\mu}  x^2 \partial_{\nu} 
\left[ i \mathbf{P}^{\nu},  z_1 \mathcal{O}_{++}^{t=2}(z_1u, \, w u) +z_2 \mathcal{O}_{++}^{t=2}(wu,  z_2 u)
\right]  \right\},
\label{eqn:app8}
\end{align}
where
\begin{align}
K^{ \mu \alpha \nu \beta }=(g^{\mu \alpha} g^{\nu \beta}-g^{\mu \nu}g^{\alpha \beta}+g^{\mu \beta} g^{\nu \alpha})-i\epsilon^{\mu \alpha \nu \beta}.
\label{eqn:app2}
\end{align}
The  twist-4 part $\mathfrak{B}_{\mu}^{t=4}(z_1, z_2)$ is more complex than the twist-2 and twist-3 ones,
\begin{align}
\mathfrak{B}_{\mu}^{t=4}(z_1, z_2)&=\frac{x^2}{8}\, \partial_{\mu}
\int_0^1  \frac{du}{u^2}
\bigg\{  u^2 (1-u^2 +u^2\ln{u})\, z_1 z_2 \mathcal{O}_{1}(z_1u, \, z_2 u)\nonumber\\
&\quad-
\left[\left( (1-u^2)\left(z_2 \partial_{z_2} -\frac{z_1}{z_{12}}\right)+(1-u^2 +u^2\ln{u}) \,z_2 \partial_{z_2}^2 z_{12}  \right)  \mathcal{R}(uz_1,uz_2) 
-(1\leftrightarrow 2)\right]
 \bigg\}.
\label{eqn:app3}
\end{align}
$\mathbb{C}(z_1, z_2)$ is pure twist 4 and can be expressed as
\begin{align}
\mathbb{C}(z_1, z_2)=&-\frac{1}{4}  \int_0^1 \frac{du}{u^2}  \,\mathcal{R}(uz_1,uz_2).
\label{eqn:app7}
\end{align}

\section{Calculation techniques for helicity amplitudes}
\label{app-b}

There are two helicity-flip amplitudes $A_{0+}$ and $A_{-+}$ in Eq.\,(\ref{eqn:am-hel-ind}). Angular momentum conservation implies that they are proportional to a given power of the transverse momentum transfer as $A_{0+} \propto \Delta \cdot \epsilon_{-}$ 
and $A_{-+} \propto (\Delta \cdot \epsilon_{-})^2$. Therefore, the twist-4 part of Eq.\,(\ref{eqn:kine-t4}) 
will be beyond the accuracy of this work due to the additional factor of 
$\Delta \cdot \epsilon_{-}$,
\begin{align}
A_{0+}=A_{0+}^{t=2}+A_{0+}^{t=3}, \qquad
A_{-+}=A_{-+}^{t=2}+A_{-+}^{t=3},
\label{eqn:t3-t2}
\end{align}
We consider the matrix  element of the twist-2 part of  $T_{0+}=T_{\mu \nu} \epsilon_0^{\mu} \tilde{\epsilon}_+^{\nu}$  and substitute it into Eq.\,(\ref{eqn:amp0}),
\begin{align}
A_{0+}^{t=2}&=\int d^4x \,e^{-ir\cdot x} 
\,\langle \bar{M}(p_2) M(p_1) | \, T_{0+}^{t=2} \,  | 0 \rangle  \nonumber\\
&=-\chi \,\frac{  \Delta \cdot \epsilon_{-} }{Q}    \int d\beta \,  d\alpha\,  \phi(\beta, \alpha) \,\beta\int_0^1 du \,\frac{4 n \cdot \tilde{n}    }{(r+ul_{z_1 z_2})^2 +i \epsilon} \nonumber \\
&= 2 \chi \,\frac{ \Delta \cdot \epsilon_{-} }{Q}
\int d\beta \, d\alpha\, \phi(\beta, \alpha)\, \beta\,  \frac{\ln(F-i\epsilon)- \ln(z_1)  }{F-z_1},
\label{eqn:amp-a01t2}
\end{align}
where
$(r+ul_{z_1 z_2})^2=-2\,n \cdot \tilde{n}\,(u F + (1-u)z_1)$ is obtained by neglecting the terms of order $\mathcal{O}(s, m^2)$ since they will not contribute at the $1/Q^2$ accuracy.
Let us mention that $i \epsilon$ can be omitted in $\ln(F-i\epsilon)$ since $F$ is always positive ($0 \le  F  \le 1$) and there is no branch cut. Indeed, $\ln(F)$  is divergent when $F=0$ $(z=1)$, but the GDA vanishes at $z=1$. Similarly, the twist-3 part is obtained,
\begin{align}
A_{0+}^{t=3}&=\int d^4x \,e^{-ir\cdot x} \,
\langle \bar{M}(p_2) M(p_1)  | \, T_{0+}^{t=3} \, | 0 \rangle  \nonumber\\
&= 2 \chi   \, \frac{ \Delta \cdot \epsilon_{-} }{Q}\,
z_2 \int d\beta\,   d\alpha \, \phi(\beta, \alpha)\,\beta
\int_0^1 du\, u \int_{z_2}^{z_1} dw \, \frac{(2 n \cdot \tilde{n})^2}{\left[ (r+ul_{w z_2})^2 +i \epsilon \right]^2 } \nonumber \\
&= 2 \chi \,\frac{ \Delta \cdot \epsilon_{-} }{Q}
 \int d\beta \,  d\alpha\, \phi(\beta, \alpha)\,  \beta\,
  \frac{-z_2 }{F-z_1} \left[ \frac{\ln(F-i\epsilon)}{F-1}  -  \frac{ \ln(z_1)  }{z_2}    \right] 
\label{eqn:amp-a01t3}
\end{align}
with 
\begin{align}
(r+ul_{w z_2})^2=-2n \cdot \tilde{n} \left[ z_1 -uw + (w-z_2) F \right].
\label{eqn:momentum1}
\end{align}
We note that although the twist-2 and twist-3 amplitudes depend on $z_1$ and $z_2$, this dependence disappears in their sum,
\begin{align}
A_{0+}=2  \chi \,  \frac{ \Delta \cdot \epsilon_{-} }{Q}
\int d\beta \,  d\alpha\, \phi(\beta, \alpha)\, \beta\,
\frac{\ln(F)  }{F-1}\,,
\label{eqn:amp-a01}
\end{align}
which indicates that  translation invariance  is recovered  in the physical amplitudes.  Once again $i \epsilon$  is  omitted in $\ln(F-i\epsilon)$ as well. 

The calculation of $A_{-+}$ is quite similar to 
the one of $A_{0+}$,
\begin{align}
A_{-+}^{t=2}&=-2 \chi\, (\Delta \cdot \epsilon_{-})^2  
\int d\beta \, d\alpha \, \phi(\beta, \alpha) \,\beta^2 
\int_0^1 du \,u \,   \frac{1}{ (r+ul_{z_1 z_2})^2 +i \epsilon } \nonumber \\
&= \chi \, \frac{(\Delta \cdot \epsilon_{-})^2 }{n \cdot\tilde{n}}  \int d\beta  \, d\alpha\, \phi(\beta, \alpha) \, \beta^2 \,\partial_F  
 \left[ \frac{F \ln(F-i\epsilon)}{F-z_1}  -  \frac{z_1 \ln(z_1)  }{F-z_1}    \right] , \nonumber \\
A_{-+}^{t=3}&=-2  \chi\, (\Delta \cdot \epsilon_{-})^2  \,n \cdot\tilde{n} 
 \int d\beta \, d\alpha\,  \phi(\beta, \alpha)\,\beta^2  \int_0^1 du\, u^2   
\int_{z_2}^{z_1} dw \left[ \frac{z_1(w-z_1)}{\left[ (r+ul_{z_1 w})^2 +i \epsilon \right]^2 } 
+\frac{z_2(z_2-w)}{\left[ (r+ul_{ wz_2})^2 +i \epsilon \right]^2 }
\right] \nonumber \\
&=-  \chi \,\frac{ (\Delta \cdot \epsilon_{-})^2}{2n \cdot\tilde{n} }   \int d\beta  \,   d\alpha\, \phi(\beta, \alpha)\,  \beta^2\,
\partial_F \left[ \frac{-\ln(F-i\epsilon)}{F-1}  +\frac{2z_1 \ln(F-i \epsilon)  }{F-z_1} 
-\frac{ 2z_1 \ln(z_1)  }{F-z_1}    \right].
\label{eqn:amp-ampt2t3}
\end{align}
For the sum, one gets 
\begin{align}
A_{-+}
= -\chi\,  \frac{(\Delta \cdot \epsilon_{-})^2}{2 n\cdot \tilde{n}}  \int d\beta  \,d\alpha \,\phi(\beta, \alpha)\,    \beta^2\,
\partial_F  \left[ \frac{1-2F}{F-1} \,\ln(F)   \right],
 \label{eqn:amp-ampsum}
\end{align}
which does not depend on $z_1$ and $z_2$,  and which is proportional to $(\Delta \cdot \epsilon_{-})^2$ as shown in
Eq.\,(\ref{eqn:am-hel-ind}).

The calculation of $A_{++}$ is more lengthy than the ones of 
helicity-flip amplitudes, since it contains the contributions of twist 2, twist 3 and twist 4.
\begin{align}
A_{++}=A_{++}^{t=2}+A_{++}^{t=3}+A_{++}^{t=4}, \qquad
A_{++}=\epsilon_{+}^{\mu} \tilde{\epsilon}_{+}^{\nu }
A_{\mu \nu}.
\label{eqn:amp-con}
\end{align}
Taking the trace of Eq.\,(\ref{eqn:amp1}), one obtains \cite{Braun:2012bg}
\begin{align}
A_{++}=-\frac{1}{2}A_{\, \  \mu}^{\mu}+ A_{\mu \nu}  \, \frac{(n^{\mu} \tilde{n}^{\nu} - \tilde{n}^{\mu} n^{\nu} )}{2 n \cdot \tilde{n} }.
 \label{eqn:amp-con-tr}
\end{align}
We simplify the operator expansion of $T_{++}$ in Eq.\,(\ref{eqn:kine-t4}) by using Eq.\,(\ref{eqn:amp-con-tr}),
and we take the matrix elements of $T_{++}$ to obtain the contribution of each twist to $A_{++}$ \cite{Braun:2012bg},
\begin{align}
A_{++}^{t=2}&=  -\int \frac{ d^4x }{\pi^2}\, \frac{e^{-ir\cdot x} }{x^4}\,
\langle \bar{M}(p_2) M(p_1)  | \, \mathcal{O}_{++}^{t=2}(z_1, \, z_2 ) \,  |0 \rangle , \nonumber\\
A_{++}^{t=3}&=2   \int \frac{ d^4x }{\pi^2}\, \frac{e^{-ir\cdot x} }{x^4} \,
(x \cdot \epsilon_{+})\,
\epsilon_{-}^{\mu}
\langle \bar{M}(p_2) M(p_1)  | \,  \mathfrak{B}_{\mu}^{t=3}(z_1, z_2)
+\mathfrak{B}_{\mu}^{t=3}(z_2, z_1) \,  |0 \rangle, \nonumber\\
A_{++}^{t=4}&=-\frac{1}{4}    \int \frac{ d^4x }{\pi^2}\, \frac{e^{-ir\cdot x} }{x^2} \int_0^1 du\,
\langle \bar{M}(p_2) M(p_1)  |  \left\{  z_1 z_2 u^2 \mathcal{O}_1(z_1u, z_2u)  \right. \nonumber \\
&\quad\left.  - 
[z_2 \partial_2^2 z_{12} \mathcal{R}(z_1u,z_2u)-(1\leftrightarrow 2)]  -2\, \mathcal{R}(z_2,z_1)   \right\}   |0 \rangle, 
\label{eqn:a11-t2t3t4}
\end{align}
where the expressions for $\mathfrak{B}_{\mu}^{t=3}(z_1, z_2)$  and  $\mathcal{R}(z_2,z_1)$ can be found in Appendix \ref{app-a}.

The calculation of $A_{++}^{t=2}$ is rather straightforward with the help of
Eq.\,(\ref{eqn:dds-t4}),
\begin{align}
A_{++}^{t=2}=&2  \chi  \int d\beta \,  d\alpha \, \phi(\beta, \alpha)
\left[ \ln((r+ul_{z_1 z_2})^2 +i \epsilon) - l_{z_1 z_2}^2 \int_0^1  du\, u   \,
\frac{1}{ (r+ul_{z_1 z_2})^2 +i \epsilon  }  \right]  \nonumber \\
=&2  \chi  \int d\beta \,  d\alpha\,  \phi(\beta, \alpha) \left\{ \ln(F-i\epsilon)+ 
\frac{\beta^2 \Delta_T^2}{8n\cdot \tilde{n}}\,
\partial_F \left[z_1\,\frac{ \ln(F-i \epsilon) -\ln(z_1)   }{F-z_1} \right] \right. \nonumber \\
&+\left. \frac{s}{2n \cdot\tilde{n}} \left[  z_1 z_2 + (F-z_1)\alpha +(1-F)F  \right]
\partial_F  
 \left[ \frac{F \ln(F-i\epsilon)}{F-z_1}  -  \frac{z_1 \ln(z_1)  }{F-z_1}  \right]
\right\},
\label{eqn:a11-t2}
\end{align}
where Eq.\,(\ref{eqn:dds-sys}) is used to eliminate the irrelevant terms, and 
the first term in Eq.\,(\ref{eqn:a11-t2}) is actually the twist-2 amplitude given by Ref.\,\cite{Diehl:1998dk}.
The twist-3 contribution is expressed as 
\begin{align}
A_{++}^{t=3}=&\frac{\chi}{n \cdot \tilde{n}}  \int d\beta \,   d\alpha\,  \phi(\beta, \alpha)\,\beta
\left[ \zeta_0 s -\frac{\Delta_T^2}{4} \,\beta\, \partial_F \right] 
\frac{\ln(F-i\epsilon)}{F-1},  
\label{eqn:a11-t3}
\end{align}
and is independent of $z_1$ and $z_2$.  There are several terms contributing to the amplitude of 
$A_{++}^{t=4}$ in Eq.\,(\ref{eqn:a11-t2t3t4}), which we denote as 
\begin{align}
A_ {++}^{t=4}=A_{(1)}^{t=4}+A_{(2+3)}^{t=4}+A_{(4)}^{t=4}.
\label{eqn:a11-t4-c4}
\end{align}
$A_{(1)}^{t=4}$ is calculated by using Eq.\,(\ref{eqn:o1o2-t4}),
\begin{align}
 A_{(1)}^{t=4}=-\chi\,\frac{ z_1 z_2 s }{n\cdot \tilde{n}}  \int d\beta\,    d\alpha\,  \phi(\beta, \alpha)\,
\partial_F  
 \left[ \frac{F \ln(F-i\epsilon)}{F-z_1}  -  \frac{z_1 \ln(z_1)  }{F-z_1}  \right].
\label{eqn:a11-t4-1}
\end{align}
When calculating the other terms in $A_{++}^{t=4}$, one finds divergences in $A_{(2)}^{t=4}$ and $A_{(3)}^{t=4}$ associated with the real (onshell) photon. Here, the photon is set as an offshell one so as to regularize the divergences,
\begin{align}
q_2 &\rightarrow q_2+(\xi -z_1) (q_1+q_2), \nonumber \\
r=z_1q_1+z_2q_2 &\rightarrow r=-q_2+\xi (q_1+q_2),
\label{eqn:offshell}
\end{align}
and we take $\xi=z_1$ at the end of the calculations to obtain the final results.
As shown in Eq.\,(\ref{eqn:a11-t2t3t4}), the operator $\mathcal{R}(z_1,z_2)$ is involved in the remaining terms in $A^{++}_{t=4}$.
First, we simplify the matrix element of $\mathcal{R}(z_1,z_2)$,
\begin{align}
&\langle \bar{M}(p_2) M(p_1) | \,  \mathcal{R}(z_1,z_2)  \, |0 \rangle  \nonumber \\
&=  -2i \chi  \int_{z_2}^{z_1} dw \,\int_{z_2}^{w} \,\frac{dw_1}{z_{12}}
 \frac{w_1-z_2}{z_1-w_1} 
 \int d\beta  \,  d\alpha \, \phi(\beta, \alpha)
 \left[ \frac{s}{2}\,S_{+}+2(S_0-1)P\cdot l_{w w_1}+ i P\cdot x \,l_{w w_1}^2
 \right]  \frac{e^{-i l_{w w_1}\cdot x}}{w-w_1},
 \label{eqn:rz1z2}
\end{align}
where $z_{12}$ is kept so that $z_{12}=1$ is not a necessary condition. One can make the replacements $z_1 \rightarrow u z_1$ and $z_2 \rightarrow u z_2$ to obtain $\mathcal{R}(z_1u,z_2u)$.
We substitute the matrix elements $\mathcal{R}(z_1u,z_2u)$ and $\mathcal{R}(z_2u,z_1u)$ into 
Eq.\,(\ref{eqn:a11-t2t3t4}) and we take $\xi=z_1$,
\begin{align}
A_{(2+3)}^{t=4}&= \frac{ \chi }{2}  \int d\beta\,   d\alpha \, \phi(\beta, \alpha)
\left\{   \frac{s}{n \cdot\tilde{n}} \left[  \alpha  +(F-1)F \,\partial_F  \right] - \frac{\beta^2 \Delta_T^2}{4n \cdot\tilde{n}} \,\partial_F   \right\} \frac{1}{F-1}
\nonumber \\
&\quad\times 
\left\{ 
z_2 \left[\frac{F \ln(F-i \epsilon)}{F-z_1} -\frac{z_1}{z_2} \,\frac{F-1}{F-z_1} \, \ln(z_1) \right]
+z_1 \left[\frac{(F-1) \ln(1-F-i \epsilon)}{F+z_2} +\frac{1-F}{F+z_2} \,  \ln(z_1) \right]
\right\},
\label{eqn:a11-t4-23}
\end{align}
 Similarly, the last contribution to 
$A_ {++}^{t=4}$ is given by
\begin{align}
A_{(4)}^{t=4}=& \chi   \int d\beta \,  d\alpha\,  \phi(\beta, \alpha)
\left\{   \frac{s}{n\cdot \tilde{n}} \left[  \alpha +(F-1)F\, \partial_F  \right] - \frac{\beta^2 \Delta_T^2}{4n\cdot \tilde{n}} \,\partial_F \right\} \frac{1}{1-F}
\left[ \ln(F-i \epsilon) -\text{Li}_2(1)+  \text{Li}_2(F+i \epsilon) \right],
\label{eqn:a11-t4-4}
\end{align}
where the replacements $\alpha \rightarrow -\alpha$, $\beta \rightarrow -\beta$ and $F \rightarrow 1-F$ are used to simplify the amplitude.

Summing over all contributions leads finally to
\begin{align}
A_{++}&= A_{++}^{t=2}+ A_{++}^{t=3} +A_{(1)}^{t=4}+A_{(2+3)}^{t=4}+A_{(4)}^{t=4} \nonumber \\
&=\chi   \int d\beta\,   d\alpha\,  \phi(\beta, \alpha)
\left\{ 2 \ln(F)  +  \left[ \frac{s}{n \cdot\tilde{n}}\,(F-\alpha) + \frac{\beta^2 \Delta_T^2  }{4n\cdot \tilde{n}} \,\partial_F    \right]   \frac{1}{F-1}   \left[ \frac{\ln(F)}{2}-\text{Li}_2(1)+ \text{Li}_2(F) \right]    \right\},
\label{eqn:a11-all-final}
\end{align}
where the $z_1$ and $z_2$ dependences disappear as expected, and where $i \epsilon$ is  omitted
in  the functions of $\ln(F-i\epsilon)$ 
and $\text{Li}_2(F-i\epsilon)$.

\end{document}